\documentclass[12pt]{article}
\usepackage{amssymb}

\usepackage{amsmath}

\newcommand{\be}{\begin{equation}}
\newcommand{\ee}{\end{equation}}
\newcommand{\bea}{\begin{eqnarray}}
\newcommand{\eea}{\end{eqnarray}}

\begin{document}

\bigskip 
\begin{titlepage}

\begin{flushright}
UUITP-13/02\\ 
hep-th/0210198
\end{flushright}

\vspace{1cm}

\begin{center}
{\Large\bf Miracles and complementarity in de Sitter space \\}

\end{center}
\vspace{3mm}

\begin{center}

{\large
Ulf H.\ Danielsson{$^1$}, Daniel Domert{$^2$} and Martin Olsson{$^3$}} \\

\vspace{5mm}

Institutionen f\"or Teoretisk Fysik, Box 803, SE-751 08
Uppsala, Sweden

\vspace{5mm}

{\tt
{$^1$}ulf@teorfys.uu.se \\
{$^2$}daniel.domert@teorfys.uu.se \\
{$^3$}martin.olsson@teorfys.uu.se \\
}

\end{center}
\vspace{5mm}

\begin{center}
{\large \bf Abstract}
\end{center}
\noindent
In this paper we consider a scenario, consisting of a de Sitter phase followed by a phase described by a scale factor
$a(t)\sim t^{q}$, where $1/3<q<1$, which can be viewed as an inflationary toy model. It is argued that this scenario
naively could lead to an information paradox. We propose that the phenomenon of Poincar\'{e}
recurrences plays a crucial role in the resolution of the paradox. We also comment on the relevance of these results to inflation
and the CMBR.

\vfill
\begin{flushleft}
October 2002
\end{flushleft}
\end{titlepage}
\newpage

\section{Introduction}

\bigskip

Recently a lot of attention has been devoted to physics in a space time with
positive cosmological constant $\Lambda $, i.e. de Sitter space. There are
observational as well as theoretical reasons for this attention. From the
observational point of view, there is strong evidence in favor of an
accelerating universe, which might be due to a positive cosmological
constant \cite{Riess:1998cb,Perlmutter:1999np}. Understanding de
Sitter space is also important in inflationary scenarios, where the
possibility of transplanckian imprints in the CMBR spectra has led to
renewed interest in the field \cite{Trans,Ulf:0210}.

On the theoretical side the recent interest is partly due to the progress
made in the understanding of quantum gravity for AdS spaces with negative $%
\Lambda $ \cite{Maldacena:1998re,Gubser:1998bc,Witten:1998qj}. 
In this case holography has played a key role and the hope has
been that similar ideas should be important for de Sitter space as well. The
cosmological horizons present in de Sitter space, and the possible parallels
with black hole physics, make the problem even more interesting and
challenging.

In this paper we will focus on the problem of complementarity in de Sitter
space. Our purpose is to investigate the possibilities of an information
paradox and compare with the corresponding situation in the case of black
holes. In black hole physics the general view is that a complementarity
principle is at work implying that two observers, one travelling into a
black hole and the other remaining on the outside, have very different views
of what is going on. According to the observer staying behind, the black
hole explorer will experience temperatures approaching the Planck scale
close to the horizon. As a consequence the black hole explorer will be
completely evaporated and all information transferred into Hawking
radiation. According to the explorer herself, however, nothing peculiar
happens as she crosses the horizon. As explained in \cite{Susskind:1993if,
Susskind:1993ki,Susskind:1993mu,Susskind:1995qc} there is
no paradox since the two observers can never meet again to compare notes.
Any attempt of the observer remaining on the outside to send a signal to the
one on the inside, will necessarily make use of planckian energies and
presumably fail.

In order to probe the possibility of a similar paradox in the case of de
Sitter space, we will consider a scenario where at some point in time the de
Sitter phase is turned off and replaced by a $\Lambda =0$ phase with scale
factor $a(t)\sim t^{q}$, for $1/3<q<1$. We will refer to this latter phase
as the post-de Sitter phase. This is, in fact, nothing but an inflationary
toy model and towards the end of the paper we will make some comments
relevant to such an application. Considering this scenario one might naively
be led to a possible information paradox. The paradox is a result of
assuming that an object receding towards the de Sitter horizon of an
inertial de Sitter observer, will return its information content to the
observer in the form of de Sitter radiation. If the cosmological constant
turns off, the object itself will eventually return to the observers causal
patch, and one has the threat of an information paradox.

As we will explain, the time scale for Poincar\'{e} recurrences will play an
important role in the resolution of the paradox. This is similar to earlier
work \cite{Silv:0205}, where recurrence plays a crucial role in another
setting, namely, the tunneling from de Sitter space to flat space. In that
case, as well as in ours, the characteristic time scale for the process is
of the order the recurrence time, implying that the process is unphysical.
In fact, the time one has to wait for the information to return classically
is of the same order as for an arbitrary miracle to occur, that is, a
breakdown of the second law by chance.

The outline of this paper is as follows. In section 2 we start by reviewing
some basic classical and semi-classical properties of de Sitter space. We
also briefly comment on quantum gravity in de Sitter space and discuss
similarities and differences to the black hole case. In section 3 we
describe the paradox in more detail, comment on various possible loopholes
and make attempts to resolve it. We briefly compare the time scales involved
with those in a realistic inflationary scenario, and touch upon some
peculiarities encountered, related to de Sitter space itself. We end, in
section 4, with some conclusions.

\section{Some aspects of de Sitter space and the relation to black hole
physics}

\bigskip

\subsection{Classical de Sitter space}

de Sitter space is the maximally symmetric vacuum solution to the Einstein
equations with a positive cosmological constant $\Lambda $. One way to
realize de Sitter space is to view it as a hyperboloid embedded in ordinary
flat Minkowski space. Four dimensional de Sitter space, which is of interest
in this paper, is then described as the hypersurface
\begin{equation}
-X_{0}^{2}+X_{1}^{2}+X_{2}^{2}+X_{3}^{2}+X_{4}^{2}=\frac{3}{\Lambda }=R^{2}.
\end{equation}
There are numerous coordinate systems that can be used in discussing the
various aspects of de Sitter space (see \cite{Spra:0110} for a nice review).
In the so called static coordinates, which are useful when we want to focus
on observations made by a particular observer, the metric takes the form
\begin{equation}
ds^{2}=-\left( 1-\frac{r^{2}}{R^{2}}\right) dt^{2}+\left( 1-\frac{r^{2}}{%
R^{2}}\right) ^{-1}dr^{2}+r^{2}d\Omega ^{2}.
\end{equation}
At $r=R$ we notice the presence of a horizon, the de Sitter horizon. From
the point of view of an observer at $r=0$, the horizon acts as a one-way
membrane preventing anything that leaves through it to ever come back again,
as long as the space-time continues to be de Sitter. We can easily
generalize to a situation with a black hole in de Sitter space described by
\begin{equation}
ds^{2}=-\left( 1-\frac{2E}{r}-\frac{r^{2}}{R^{2}}\right) dt^{2}+\left( 1-%
\frac{2E}{r}-\frac{r^{2}}{R^{2}}\right) ^{-1}dr^{2}+r^{2}d\Omega ^{2}.
\end{equation}
In general this space has two horizons, the black hole and the de Sitter
horizon respectively. As $E$ is increased it can be seen that the two
horizons approach each other, and for $E=3^{-3/2}R$ the two horizons
coincide at $r=3^{-1/2}R$. As a consequence there is a maximal sized black
hole which can fit inside the de Sitter space.

In discussing cosmology, a more convenient choice of coordinates is the FRW
coordinates
\begin{equation}
ds^{2}=-dt^{2}+a(t)^{2}(dr^{2}+r^{2}d\Omega ^{2}),
\end{equation}
where $a(t)=Re^{t/R}$, and the de Sitter radius $R$ is related to the Hubble
constant, $H$, through $R=1/H$. With these coordinates, covering half of de
Sitter space, one can address global questions such as those relevant for a
cosmology with a period of inflation.

\subsection{Semiclassical de Sitter space}

\label{sec:semi} Gibbons and Hawking \cite{GibbHawk:1977} noted that the de
Sitter horizon possesses surface gravity and obeys laws analogous to those
that govern the physics of black holes. This analogy suggests that the de
Sitter horizon can be associated with a temperature and an entropy, similar
to what is the case for the black hole event horizon. A black hole steadily
emits Hawking radiation, which can be interpreted as the spontaneous
creation of particles at a point just outside the black hole horizon, and
one would expect that the same kind of radiation would be emitted from the
de Sitter horizon. Gibbons and Hawking demonstrated that an observer in de
Sitter space indeed detects thermal radiation at a temperature
\begin{equation}
T_{dS}=\frac{1}{2\pi R},
\end{equation}
in planckian units, and that the dS horizon can be endowed with an entropy
\begin{equation}
S_{dS}=\pi R^{2}=\frac{3\pi }{\Lambda },
\end{equation}
supporting the analogy between the de Sitter horizon and the event horizon
of a black hole.

Pointing out these similarities, there are some important differences
between the two cases worth mentioning. The black hole horizon is an
observer independent construction, while the de Sitter horizon is an
observer dependent. Any observer in de Sitter space is surrounded by a de
Sitter horizon, and if the observer moves, the horizon does so as well. So
every observer in de Sitter space lives in the center of a ``bubble'',
bounded by an event horizon with the radius $R$. As a consequence, the
relation between different observers in de Sitter space is, unlike the black
hole case, symmetric.

Another crucial difference is that de Sitter space usually is assumed to be
in thermal equilibrium.\footnote{%
An exception is the work \cite{Mottola:ar,Mottola:qt} where the
cosmological constant is claimed to relax due to the radiation.} The horizon
not only emits radiation, it also absorbs radiation, previously emitted by
itself, at the same rate, keeping the radius fixed (under the assumption
that $\Lambda $ is kept fixed). In the case of black holes, on the other
hand, one has the option of either studying black holes in thermal
equilibrium with a heat bath, or black holes that are truly evaporating.

\subsection{Quantum gravity in de Sitter space?}

\label{sec:qgr} Lacking a true quantum gravity description of de Sitter
space, one can only speculate on what kind of features it would possess.
Many believe that the describing theory should be holographic. That is, the
fundamental degrees of freedom should be the boundary degrees of freedom,
with no more than one degree of freedom per Planck area. For a discussion
of this interesting idea, see \cite{Banks:0007,Witten:2001kn}. This gives
particularly strong constraints for de Sitter spaces, simply because the
operationally meaningful part of de Sitter space is bounded by the (finite
sized) de Sitter horizon. This implies that a microscopic description of de
Sitter space should only have a finite number of degrees of freedom, i.e.
the entropy $S_{dS}=\pi R^{2}$ should better be thought of as the total
number of degrees of freedom describing the universe. Microscopically
deriving this number is a great challenge.\footnote{%
There are, however, suggestions that this number is a defining property of
the theory rather than some derivable consequence of it \cite{Banks:0007},
i.e. in order to end up with a theory with a finite number of degrees of
freedom one should postulate it from the beginning rather than deriving it
in the end.}

Much of this parallels the situation with black holes, where, indeed, the
entropy has been microscopically derived in some special cases, for early
references see  \cite{Vafa:9601,Callan:1996dv,Das:1996wn}. A
microscopic description of black holes implies that they are not as
featureless as the classical and semiclassical description suggests, and, in
particular, the radiation should not be expected to be purely thermal, but
rather be able to carry information. It is reasonable to assume that the
same is true for the de Sitter radiation, even though it should be noted
that radiation carrying information not necessarily implies that an observer
will be able to extract information from it. We will return to this
important point later on.

\section{The paradox and possible resolutions}

\bigskip

\subsection{The scenario}

\label{sec:scen} As mentioned in the introduction, our scenario can globally
be viewed as an inflationary toy model. The universe starts out in a de
Sitter phase with a scale factor $a(t)\sim e^{t/R}$. At a certain point in
time, $t_{0}$, the de Sitter phase is turned off and replaced by a post-de
Sitter phase where the universe is filled with matter with an equation of
state given by $p=\sigma \rho $, and a scale factor $a(t)\sim t^{q}$ with $q=%
\frac{2}{3\left( 1+\sigma \right) }$. We will consider values of $q$ in the
range $1/3<q<1$, where the lower bound comes from the requirement that $%
\left| \sigma \right| \leq 1$, and the upper bound is the condition for not
having an accelerating universe, which would prevent our paradox from being
realized.

At the time $t_{0}$ of transition we would like to match the scale factor $%
a(t)$ and the Hubble constant $H=\dot{a}/a$ smoothly. We imagine that the
cosmological constant rapidly decays and the energy is transferred into
matter through reheating. In the de Sitter phase we have
\begin{equation}
a(t)=R e^{t/R}, \hbox{ }H=\frac{1}{R},
\end{equation}
and in the post-de Sitter phase we make the following ansatz for the scale
factor 
\begin{equation}
a(t)=A(t-B)^{q}.
\end{equation}
The continuity of $a(t)$ and $H$ at $t=t_{0}$ requires that 
\begin{equation}
A=\frac{Re^{t_{0}/R}}{(qR)^{q}}
\end{equation}
\begin{equation}
B=t_{0}-qR,
\end{equation}
which gives us the scale factor during the post-de Sitter phase according to 
\begin{equation}
a(t)=\frac{Re^{t_{0}/R}}{(qR)^{q}}(t-t_{0}+qR)^{q}.
\end{equation}
After these introductory remarks we will move to the possibility of
obtaining a paradox.

\subsection{An information paradox?}

\label{sec:par} Let us consider an inertial observer during the de Sitter
phase, who drops an object that recedes towards the horizon. Taking the
perspective of global coordinates, the object will eventually leave the
causal patch of the observer and information is apparently lost. However, as
mentioned in section \ref{sec:qgr}, the radiation coming from the horizon is
expected, as in the case of black holes, to carry information. By itself,
this gives rise to some highly non-trivial questions and issues, discussed
in section \ref{sec:rinf}.

In our scenario, this is however not the end of the story. By turning off
the de Sitter phase, a new possibility for the information to get back
emerges, namely, the object itself can return to within the causal patch of
our observer, confusing her by apparently duplicating the information.

One natural question at this stage would be how long time it takes for the
object to get back inside the causal patch of the observer. A short
calculation gives us the answer. The condition for the return of the object
to the observers causal region is that the distance, in global coordinates,
between the observer and the object should be equal to the causal size of
the observers part of the universe. Since the object moves away from the
observer at the rate of the expansion of the universe, i.e. as $\sim t^{q}$
with $q<1$, and the observers light cone grows like $\sim t$, one concludes
that at some point in time the two distances will coincide. For concreteness
we will consider an object at constant comoving coordinate $x$ with $%
x\lesssim 1$. At time $t=0$ the object is consequently at a physical
distance $xR\lesssim R$ from the observer. The physical distance between the
object and the observer at the time of return, $t_{in}$, is given by 
\begin{equation}
xR\mbox{ }e^{\frac{t_{0}}{R}}\left( \frac{a(t_{in})}{a(t_{0})}\right) =xR%
\mbox{ }e^{\frac{t_{0}}{R}}\left( \frac{t_{in}-t_{0}+qR}{qR}\right) ^{q}.
\label{eq:odist}
\end{equation}
The causal size of the universe for the observer at $t=t_{in}$ is 
\begin{equation}
\begin{split}
& (t_{in}-t_{0}+qR)^{q}\int_{t_{0}}^{t_{in}}\frac{dt}{(t-t_{0}+qR)^{q}} \\
& =\frac{1}{1-q}(t_{in}-t_{0}+qR)^{q}\left(
(t_{in}-t_{0}+qR)^{1-q}-(qR)^{1-q}\right) .
\end{split}
\label{eq:cdist}
\end{equation}
Identifying (\ref{eq:odist}) and (\ref{eq:cdist}) and solving for $t_{in}$
gives
\begin{equation}
 \label{eq:tin1}
t_{in}=\left( \mbox{ }\frac{(1-q)xR\mbox{ }e^{\frac{t_{0}}{R}}}{(qR)^{q}}%
+(qR)^{1-q}\right) ^{\frac{1}{1-q}}+t_{0}-qR.
\end{equation}
As a consequence it seems like the information has returned twice to our
observer, under the assumption that the observer can get complete
information from the de Sitter radiation. In the following two subsections
we discuss possible loopholes in the arguments leading to this apparent
puzzle.

Before doing this, however, it is illuminating to consider what this would
look like locally, i.e. what our observer actually would see happen. Let us
assume that the object is continuously emitting signals towards our
observer. As the object approaches the de Sitter horizon, the signals become
increasingly redshifted and effectively disappears from the observers sight.
As long as we are in the de Sitter phase, the redshift increases
exponentially. In the post-de Sitter phase, however, we have a redshift that
steadily decreases, meaning that the object eventually will become visible
to the observer at a certain time in the distant future. Let us calculate
this time and compare it with the time $t_{in}$ derived above.

For simplicity we focus on the case where not only the time of detection of
the signal, $t_{obs}$, but also the time of emission, $t_{em}$, are in the
post-de Sitter phase. We then find that
\begin{equation}
\begin{split}
x& =\int_{t_{em}}^{t_{obs}}\frac{dt}{a(t)}=\frac{1}{A}\left[ \frac{1}{1-q}%
(t-B)^{1-q}\right] _{t_{em}}^{t_{obs}} \\
& =\frac{1}{(1-q)A}\left( (t_{obs}-B)^{1-q}-(t_{em}-B)^{1-q}\right) ,
\end{split}
\end{equation}
where $A$ and $B$ was calculated in section \ref{sec:scen}. Using this, the
redshift, $z$, becomes
\begin{equation}
1+z=\frac{a(t_{obs})}{a(t_{em})}=\frac{(t_{obs}-B)^{q}}{(t_{em}-B)^{q}}=%
\frac{(t_{obs}-B)^{q}}{\left( (t_{obs}-B)^{1-q}-Ax\left( 1-q\right) \right)
^{\frac{q}{1-q}}},  \label{eq:red}
\end{equation}
where it can be noted that for $t_{obs}\rightarrow \infty $ the expression
goes to unity as expected. The condition for the object to become visible
again is
\begin{equation}
(t_{obs}-B)^{1-q}\gtrsim Ax\left( 1-q\right) ,
\end{equation}
rendering (\ref{eq:red}) of order one. Solving for $t_{obs}$, we get
\begin{equation}
t_{obs}\gtrsim \left( Ax\left( 1-q\right) \right) ^{\frac{1}{1-q}}+B=\left(
\frac{(1-q)xR}{(qR)^{q}}\right) ^{\frac{1}{1-q}}e^{\frac{t_{0}}{(1-q)R}%
}+t_{0}-qR,
\end{equation}
which indeed is of the same order as $t_{in}$, given by (\ref{eq:tin1}).

\subsection{Measuring information in the de Sitter radiation}

\label{sec:rinf} In the last subsection, a crucial point for the occurrence
of the paradox was that one can actually extract information from the de
Sitter radiation. Is this really possible? If so, how much radiation is
needed and how long time will it take to receive the appropriate amount?

In order to answer these questions, one would need a full quantum analysis
of the process. Following a less ambitious route, where one considers some
very general entropy relations, one can at least obtain a reasonable
estimate of the quantities involved. This was done by Page in \cite
{Page:9305}, where the main idea was to consider a total system, in a random
pure state, being built up by two subsystems. The question then was how much
information could be expected to be contained in the different parts, i.e.
the smaller subsystem, the larger subsystem and the correlations between the
two. In our situation this would translate into asking how much information
we have in the horizon, in the radiation and in the correlations between
them. Page's calculations revealed that the information in the smaller
subsystem is always less than one half bit of information, thus basically
containing no information at all.

Let us now consider a measuring process where a detector collects radiation
and, perhaps, information. As we saw from the discussion of the
Schwarzschild-de Sitter metric in section \ref{sec:semi} our detector must
be limited by the size of the largest black hole that can fit in de Sitter
space. This means that it can never access more than one third of the
degrees of freedom (as was also discussed in \cite{Bousso:0205}).
The argument of Page would furthermore suggest that
almost no information will be located in the detector but rather it resides
in the correlations.\footnote{%
Even if this is correct, this does not necessarily imply that the situation
is clear cut. We still have to consider the possibility of extracting
quantum information in the post-de Sitter phase, where the state of the
detector must be correlated with something replacing the de Sitter horizon.
One possibility is that the state of the detector is correlated with the
matter created through reheating. Then it may, at least in principle, be
possible to eventually extract the quantum information.}

This discussion crucially hinges on the applicability of Page's argument to
the discussion of complementarity. In his calculation it is assumed that one
is considering a \textit{random} pure state of the total system. In our case
it is not clear that this is a fair description. After all, a detector (and
observer) prepared to take part in a measuring process is not just any
state. But, as we will show, even if the information \textit{can} be
retrieved in the de Sitter phase, this will not necessarily lead to a
paradox anyway.

Now, what is the minimum time needed to measure the information in the
radiation? A reasonable estimate would be the time it takes for the $R^{2}$
degrees of freedom of the horizon to, at least in principle, become
available to our observer. One would expect this to be the time it takes for
an amount of entropy (or information) of the order of $R^{2}$ to circulate
once through the system. That is, being emitted by the horizon and
reabsorbed again. The total flow of entropy (per unit time) from the horizon
is given by $T^{3}R^{2}$, and the time to transfer $R^{2}$ is then $\tau
\sim R^{2}/\left( T^{3}R^{2}\right) \sim T^{-3}\sim R^{3}$, where we have
used that $T\sim \frac{1}{R}$.

One can argue for the same result in the following illuminating way. A
single particle in de Sitter space would need roughly the time $R$ to pass
through the causal patch of the observer. We then need to know how many
particles there are in the de Sitter radiation at a given time. A rough
calculation uses the fact that the number density, $n$, of black body
radiation at the temperature $T$, goes like $n\sim T^{3}$. This means that
the total number of particles, $N$, in a horizon volume $V\sim R^{3}$
becomes
\begin{equation}
N=nV\sim \frac{1}{R^{3}}R^{3}=1.
\end{equation}
This result\footnote{%
The fact that there is just one particle per horizon volume was also pointed
out in \cite{Banks:0102}, but has in general not received a lot of attention.%
} might seem somewhat surprising and is not really compatible with what
seems to be the general picture of an observer bathing in a sea of de Sitter
radiation. Now it is straightforward to estimate the time it takes for $%
R^{2} $ particles to, at least in principle, become available. Since one
particle needs the time $t\sim R$ and we have roughly one particle per
horizon volume, we recover $\tau \sim R^{3}$ in the case of $R^{2}$ particles.
This estimate will be used in the following subsection.

\subsection{Exceeding the recurrence time?}

In section \ref{sec:par} we calculated the time it takes for the object to
be available again for the observer. The result, (\ref{eq:tin1}), has a
dependence on $t_{0}$ which is the time passed from the moment the object is
released to the time the de Sitter phase is turned off. Then it is natural
to let this time be at least the time $\tau $ estimated in the last
subsection, i.e. the time needed for the observer to, at least in principle,
be able to collect enough radiation during the de Sitter phase.

Now, since $\tau $ is a large time, being the cube of $R$, we only consider
the dominant term in (\ref{eq:tin1}). Hence the time of return is
\begin{equation}
t_{in}\sim e^{\frac{1}{1-q}\frac{\tau }{R}}\sim e^{R^{2}}\sim e^{S_{dS}},
\end{equation}
where it is used that the entropy in de Sitter space is $S_{dS}=\pi R^{2}$,
up to factors of order one in the exponential. But this is nothing but the
Poincar\'{e} recurrence time for our de Sitter space! The recurrence time is
the time it takes for a trajectory in phase space (for an isolated finite
system) to return arbitrarily close to its initial value. In particular,
this means that discussing experiments lasting longer than their recurrence
time is meaningless, since the memory of the system then has been erased.
Since the detector obviously has a smaller entropy than the entire de Sitter
space, this suggests that we are considering an experiment lasting for too
long to make any sense. Thus if the detector can be considered as an
isolated system by itself in the post-de Sitter phase (until the expected
signal is coming back), the paradox is eliminated. On the other hand, if the
detector is allowed to interact with the environment, with more degrees of
freedom coming into play, the relevance of Poincar\'{e} recurrences is not
clear. At the same time, however, the issue of unitarity and a possible
information paradox comes in a different light since we are considering an
open system.

A related discussion regarding non-unitary processes can be found in \cite
{Silv:0205}, where tunneling from de Sitter to flat space was considered. In
that case the entire causal diamond disappears, corresponding to a severe
violation of unitarity. However, it was argued that this process should be
considered unphysical, since the time needed for the process to occur was of
the order of the recurrence time for the de Sitter space. This is in the
same spirit as the suggested resolution of our paradox, as discussed above.

To summarize, if we consider a detector in de Sitter space, there does not
seem to exist a physical process such that information beyond the horizon
can be made accessible to the detector before the recurrence time of the
detector has passed. This includes tunneling processes as discussed in \cite
{Silv:0205}, as well as an abrupt end of inflation as discussed in this
paper. In other words, the observer needs to wait as long as the time needed
for a thermodynamical miracle to happen.

\subsection{Comparisons with a realistic inflationary cosmology}

At this point it is interesting to compare our results with the time scales
involved in a realistic inflationary cosmology, just to give a taste of the
relevant order of magnitudes. To solve the classical cosmological problems,
we need the number of e-foldings to be at least 70 or so, corresponding to a
time $t_{infl}$, roughly given by $t_{infl}\sim 70/H$, if we consider $H$ as
time-independent. The value of $H$ is constrained from observations of the
CMBR to be no larger than $10^{-4}\mbox{ }m_{Pl}$, which corresponds to $%
t_{infl}\sim 7\cdot 10^{5}\mbox{ }t_{Pl}$. The fluctuations that eventually
will be visible in the CMBR emerges out of planckian scales well within this
time frame, counted backwards from the end of inflation. Clearly this is
much shorter than
\begin{equation}
\tau \sim R^{3}=1/H^{3}\sim 10^{12}\mbox{ }t_{Pl},
\end{equation}
and as a consequence the issue of complementarity would not be expected to
be relevant. Another way to say the same thing, is to estimate the time $%
t_{in}$ needed for an object, that left the time $\tau $ before the end of
inflation, to reenter the causal patch of the observer. This is given by
\begin{equation}
t_{in}\sim e^{R^{2}}=e^{\frac{1}{H^{2}}}\sim e^{10^{8}}\mbox{ }t_{Pl},
\end{equation}
which exceeds the present age of the universe by many orders of magnitude.

It follows that a detector present during inflation, when the fluctuations
relevant for the CMBR are generated and leave the de Sitter horizon, can 
survive long enough to
detect their return in the subsequent post-de Sitter era. One concludes,
from the point of view of the detector, that these fluctuations will not
have time to holographically thermalize on the horizon before the end of
inflation. In other words, an observer present during the de Sitter phase
will not have time to study the de Sitter radiation long enough, before the
end of inflation and after the fluctuations are generated, to encounter any
problems with respect to information.

On the other hand, as pointed out in \cite{Ulf:0210}, as far as the CMBR
fluctuations are concerned we are more in the position of an observer inside
a black hole and, as a consequence, a different outcome of the above
calculation would not have been of great relevance.

\section{Conclusions}

\bigskip

In this paper, we have discussed the possibility of an information paradox,
arising in a specific cosmological scenario. The scenario resembles an
idealized inflationary scenario, where we go from a pure de Sitter phase to
a phase described by a scale factor $a(t)\sim t^{q}$, where $1/3<q<1$. This
could possibly lead to a duplication of information, by allowing an
observer to receive information about an object, previously disappeared
through the horizon, in two different ways. First, she could extract the
information from the de Sitter radiation, assuming this could be done.
Secondly, the object itself could return to the observer, well after the
time of transition between the two different phases involved.

Assuming that it is possible to retrieve information during the de Sitter
phase, we estimated the time needed to collect the appropriate amount of
radiation, which turned out to be $\tau \sim R^{3}$. We then turned to a
calculation of the time it would take for the object to return to the
observer, thereby apparently duplicating the information. This time turned
out to be dominated by a factor $\sim e^{R^{2}}$, using $\tau $ as input
data. The appearance of this factor suggested that a recurrence argument
could be invoked, related to issues discussed in \cite{Silv:0205}. As a
consequence, the observer will have to wait as long for the classical return
of the information as she must wait for a miracle.
\\ \\
{\bf Note added:} When this work was being completed we received \cite{Banks:0210}
that discusses the measurement problem in de Sitter space.

\section*{Acknowledgments}

UD is a Royal Swedish Academy of Sciences Research Fellow supported by a
grant from the Knut and Alice Wallenberg Foundation. The work was also
supported by the Swedish Research Council (VR).

\end{document}